\documentclass[%
 reprint,
 amsmath,amssymb,
 aps,
pra,
]{revtex4-1}
\usepackage{graphicx}% Include figure files
\usepackage{dcolumn}% Align table columns on decimal point
\usepackage{bm}%		
\begin{document}

\preprint{APS/123-QED}

\title{The influence of the enhanced vector meson sector
 on the properties\\ of the matter of neutron stars.}%with Forced Linebreak}% Force line breaks with \\

\author{Ilona Bednarek}
% \altaffiliation[Also at ]{}%Lines break automatically or can be forced with \\
\author{Ryszard Manka}%
\altaffiliation[Retired]{}%Lines break automatically or can be forced with \\
\author{Monika Pienkos}
\affiliation{%
 Department of Astrophysics and Cosmology, Institute of Physics, 
 University of Silesia, Uniwersytecka 4, 40-007 Katowice, Poland
}%


%\affiliation{
%\affiliation{
%\affiliation{%


\date{\today}% It is always \today, today,
\begin{abstract}
This paper gives an overview  of the  model of a neutron star with non-zero strangeness constructed within the framework of the nonlinear realization of the chiral $SU(3)_{L}\times SU(3)_{R}$ symmetry. The emphasis is put on the physical properties of the matter of a neutron star.
The obtained solution is particularly  aimed at the problem of the construction of a theoretical model of a neutron star matter with hyperons that will give high value of the maximum mass.

\begin{description}
\item[PACS numbers]
97.60.Jd, 26.60.Kp, 21.65.Mn, 14.20.Jn
%use the optional argument of the
%\verb+
%\item[ command] to give the category of each item.
\end{description}
\end{abstract}

\pacs{Valid PACS appear here}% PACS, the Physics and Astronomy
\maketitle
%%%%%%%%%%%%%%%%%%%%%%%%%%%%%%%%%%%%%%%%%%%%%%%%%%%%%%%%%%%%%%%%%%%%%%
\section{Introduction}

The description of the core of a neutron star is modelled on the basis of the equation of state (EoS) of dense asymmetric nuclear matter \cite{UstronEOS}. The matter density ranges from a few times the saturation density ($n_{0}$) to about an order of a magnitude higher at the core of a neutron star and at such densities hyperons are expected to emerge \cite{hypstar}. The purpose of this paper is to study the impact of hyperons on the properties of the matter of neutron stars, and on their structure. In general, analysis of the role of strangeness in nuclear structure in the aspect of multi-strange system is of great importance for both nuclear physics and for astrophysics and leads to a proper understanding  of the properties of a hyperon star.

The relativistic approach to the description of nuclear matter developed by Walecka \cite{Walecka2004} is very successful in describing 
a variety of the ground state properties of finite nuclei. Although the original Walecka  model properly describes the saturation point and the data for finite nuclei, it has been insufficient to properly describe the compression modulus of symmetric nuclear matter at saturation density.
The nonlinear self-interactions of the scalar field (the cubic and quartic terms) were added in order to get an acceptable value
of the compression modulus \cite{Boguta1977,Bodmer1989}. 
Additionally the inclusion of a quartic vector self-interaction term softens the high density component of the EoS \cite{Sugahara1994}.
The estimation of the incompressibility coefficient of symmetric nuclear matter $K_{0}$, which is 
made on the basis of recent experimental data, points to the range $240\pm 10$
MeV \cite{Piekarewicz2009}.

Models that satisfactorily reproduce the saturation properties of symmetric nuclear matter lead to considerable differences in a case in which  asymmetry dependence is included \cite{Furnstahl1996, Furnstahl1997}. Thus, the proper model of the matter of  neutron stars requires
taking  the effect of neutron-proton asymmetry into consideration. This, in turn, leads to the inclusion of the isovector meson $\rho$.
The standard version of the introduction of the $\rho$ meson field is of a minimal type without any nonlinearities. This case has been further
enlarged by the nonlinear mixed isoscalar-isovector couplings, which modify the density dependence of the $\rho$ mean field and the symmetry
energy \cite{rmib, ph, Singh2013}. The analysis of the nonlinear models should include results obtained for the relativistic FSUGold parametrisation
\cite{FSUGold}. 

A theoretical description of strangeness-rich nuclear matter requires the extension of the model to the full octet of baryons and additional meson fields were introduced to reproduce the hyperon--hyperon interaction.The model that is considered is constructed on the basis of the  hadronic $SU(3)$ theory, which naturally includes nonlinear scalar and vector interaction terms. The characteristic feature of the model is the very special form of the vector meson sector, which permits more accurate description of asymmetric strangeness-rich neutron star matter \cite{twomass}. The primary goal of this paper is to maximize the understanding of the influence of the nonlinear vector meson couplings on the form of the EoS and through this on a neutron star structure.

The calculated EoS for the matter of neutron stars in the case when hyperons are included have shown considerable stiffening for higher densities. Having obtained the EoS the analysis of the maximum achievable neutron star mass for a given class of models can be performed. Observational results limit the value of a neutron star mass and thereby put constraints on the EoS of high density nuclear matter. Recent observations  point to the existence of a high maximum neutron star mass \cite{Demorest2010,Antoniadis2013}, what is inconsistent with  theoretical models that involve hyperons.
%%%%%%%%%%%%%%%%%%%%%%%%%%%%%%%%%%%%%%%%%%%%%%%%%%%%%%%%%%%%%%%%%%%%%%
\section{The model}
%%%%%%%%%%%%%%%%%%%%%%%%%%%%%%%%%%%%%%%%%%%%%%%%%%%%%%%%%%%%%%%%%%%%%%
Recent observations of the binary millisecond pulsar J1614-2230 \cite{Demorest2010} and J0348+0432 \cite{Antoniadis2013} have led to an estimation of the neutron star mass within the range respectively $(1.97\pm 0.04) M_{\odot}$ and $(2.01\pm 0.04) M_{\odot}$. This places the maximum neutron star mass at rather high values and rules out most of the EoSs with hyperons as models that involve exotic particles predict maximum neutron star masses well below the stated value. There is a need to analyse whether it is possible to construct an EoS of neutron star matter that gives adequately high maximum mass despite including hyperons.
In this paper a description of nuclear matter based on an effective model constructed within the framework of the nonlinear realization of the chiral $SU(3)_{L}\times SU(3)_{R}$ symmetry. Details can be found in the papers by Papazoglou et al. \cite{Papazoglou1998,Papazoglou1999}.

Baryons and mesons constitute the basic degrees of freedom of the model, and consequently the Lagrangian density function  $\mathcal{L}$ splits into parts that are adequate to describe baryon $\mathcal{L}_{\mathcal{B}}$ and meson $\mathcal{L}_{\mathcal{M}}$ sectors supplemented by the
term that represents baryon--meson interactions $\mathcal{L}_{int}$, and takes the form $\mathcal{L}=\mathcal{L}_{\mathcal{B}}+\mathcal{L}_{\mathcal{M}} + \mathcal{L}_{int}$.

The meson sector of the considered model includes spin zero and spin one meson states. Nonets of different meson types, spin zero (scalar) and spin one (vector), can be written as the sum of the singlet and octet matrixes $\mathcal{\mathcal{M}}=\mathcal{M}_{sin}+\mathcal{M}_{oct}$.
Under the assumption of SU(3) symmetry, a very general form of the interaction Lagrangian $\mathcal{L}_{int}$ includes a mixture of the symmetric ($D$-type) and antysymmetric ($F$-type) couplings and  the $S$-type coupling that denotes the meson singlet state interaction
%%%%%%%%%%%
\begin{widetext}
\begin{eqnarray}
  \label{int}
\mathcal{L}_{int} & = & -\sqrt{2}g^{\mathcal{M}}_{8}(\alpha_{\mathcal{M}} [\bar{\mathcal{B}}\mathcal{B}\mathcal{M}]_{F}+(1-\alpha_{\mathcal{M}})[\bar{\mathcal{B}}\mathcal{B}\mathcal{M}]_{D})-
 g_{1}^{\mathcal{M}}\frac{1}{\sqrt{3}}[\bar{\mathcal{B}}\mathcal{B}\mathcal{M}]_{sin} =\\ \nonumber
  &=& -\sqrt{2}g^{\mathcal{M}}_{8}[\alpha_{\mathcal{M}}(Tr(\bar{\mathcal{B}}\mathcal{M}_{oct}\mathcal{B})-Tr(\bar{\mathcal{B}}\mathcal{B}\mathcal{M}_{oct}))+(1-\alpha_{\mathcal{M}})(Tr(\bar{\mathcal{B}}\mathcal{M}_{oct}\mathcal{B})+Tr(\bar{\mathcal{B}}\mathcal{B}\mathcal{M}_{oct}))]-\\  \nonumber
  &-&g^{\mathcal{M}}_{1}\frac{1}{\sqrt{3}}Tr(\bar{\mathcal{B}}\mathcal{B})Tr(\mathcal{M}_{sin}),
  \end{eqnarray}
  \end{widetext}
%%%%%%%%%%%
where the explicitly given baryon matrix $\mathcal{B}$ has the form
%%%%%%%%%%%
\begin{equation}
\mathcal{B}=\left(\begin{array}{ccc}
\frac{1}{\sqrt{6}}\Lambda+\frac{1}{\sqrt{2}}\Sigma^{0} & \Sigma^{+} & p\\
\Sigma^{-} & \frac{1}{\sqrt{6}}\Lambda-\frac{1}{\sqrt{2}}\Sigma^{0} & n\\
\Xi^{-} & \Xi^{0} &
-\frac{2}{\sqrt{6}}\Lambda\end{array}\right).
\end{equation}
%%%%%%%%%%%
Generally, the baryon--meson interaction is characterised by the following coupling constants:
the octet $g_{8}^{\mathcal{M}}$ and singlet $g_{1}^{\mathcal{M}}$ coupling constant, the parameter $\alpha_{\mathcal{M}}$, which stands for the $F/(F+D)$ ratio, and a mixing angle $\theta_{\mathcal{M}}$  that relates the physical meson fields to the pure octet and singlet states. The index $\mathcal{M} $ concerns the interaction of baryons with scalar ($\mathcal{M} = S$) and vector ($\mathcal{M} = V$) mesons (pseudoscalar and axial vector mesons, which have a vanishing expectation value at the mean field level, are not considered in this model).
In the case of the scalar meson sector baryon masses are generated by the vacuum expectation value that is attained by two scalar meson fields 
and the parameters $g_{1}^{S}, g_{8}^{S}$, and $\alpha_{S}$ have been chosen to fit the experimental values of the baryon-octet masses \cite{Papazoglou1998,Papazoglou1999}. Considering the vector meson sector of this model the octet matrix
%%%%%%%%%%%
\begin{equation}
V_{oct}=\left(
\begin{array}{ccc}
\frac{v_{8}}{\sqrt{6}}+\frac{\rho^{0}}{\sqrt{2}} & \rho^{+} & K^{\ast+}\\
\rho^{-} & \frac{v_{8}}{\sqrt{6}}-\frac{\rho^{0}}{\sqrt{2}} & K^{\ast0}\\
K^{\ast-} & \overline{K}^{\ast0} & -\frac{2v_{8}}{\sqrt{6}}\end{array} 
\right)\label{VM}
\end{equation}
%%%%%%%%%%%
supplemented with the singlet state $$V_{sin}= \frac{1}{\sqrt{3}}diag(v_{0},v_{0},v_{0})$$ comprises the vector meson nonet.
The combinations of the unphysical SU(3) singlet ($v_{0}$) and octet ($v_{8}$) states produce the physical $\omega$ and $\phi$ mesons
%%%%%%%%%%% 
\begin{eqnarray}
\phi&=&v_{8}\cos{\theta_{V}}-v_{0}\sin{\theta_{V}},\\ \nonumber
\omega&=&v_{8}\sin{\theta_{V}}+v_{0}\cos{\theta_{V}}.
\end{eqnarray} 
%%%%%%%%%%%
The $\phi$ meson  taken as pure $\bar{s}s$ state leads to the ideal mixing with $\theta_{V} \approx 35.3^{\circ}$.
If the determination of the baryon--vector meson couplings bases on the assumption that nucleons do not couple to the $\bar{s}s$ meson, then $g_{1}^{V}=\sqrt{6}g_{8}^{V}$.
In the limit $\alpha_{V} = 0$ (only the $F$-type coupling remains), the coupling constants are related to the additive quark model and the
vector meson coupling constants are given by the following relations:
%%%%%%%%%%%
\begin{eqnarray}
\label{vectorcouplings}
g_{\Lambda\omega}&=&g_{\Sigma\omega}= 2g_{\Xi\omega}=\frac{2}{3}g_{N\omega}=2g_{8}^{V},\\ \nonumber
g_{\Lambda\phi}&=&g_{\Sigma\phi}=\frac{g_{\Xi\phi}}{2}=\frac{\sqrt{2}}{3}g_{N\omega}.
\end{eqnarray}
%%%%%%%%%%%
The high density limit of the EoS of dense matter in neutron star interiors is dominated by the  contribution that come from the baryon number density, which justifies the construction of a model that includes a broad spectrum of  mixed vector meson couplings.
The extended vector meson sector, which stems from the SU(3) invariants, can be written in the following form
%%%%%%%%%%%
\begin{equation*}
\mathcal{L}_{V}=\frac{1}{4}c\,(Tr(VV))^{2}+\frac{1}{2}d\,
Tr((VV)^{2})+\frac{1}{16}f(Tr(V))^{4},
\end{equation*}
%%%%%%%%%%%
where $V$ represents the matrix of the vector meson fields, and $c, d, f$ are the general coefficients that have been determined by assuming that in the case of a neutron star with zero strangeness, the model described by the TM1 parameter set \cite{Sugahara1994} is recovered.
This assumption leads to the Lagrangian function, which embodies contributions from the baryon and meson sectors supplemented by the parts that describe the baryon and meson interactions
%%%%%%%%%%%
\begin{widetext}
%%%%%%%%%%%
\begin{eqnarray}
\label{lag1}
{\mathcal{L}}
&=&\sum_{B}\overline{\psi}_{B}\left(i\gamma^{\mu}D_{\mu}-m_{eff,B}\right)\psi_{B}
+\frac{1}{2}\partial^{\mu}\sigma\partial_{\mu}\sigma 
-\frac{1}{2}m_{\sigma}^{2}\sigma^{2}
-\frac{1}{3}g_{3}\sigma^{3}
-\frac{1}{4}g_{4}\sigma^{4}
+\frac{1}{2}\partial^{\mu}\sigma^{\ast}\partial_{\mu}\sigma^{\ast}
-\frac{1}{2}m_{\sigma^{\ast}}^{2}\sigma^{\ast 2}
+\\ \nonumber
&+&\frac{1}{2}m_{\omega}^{2}(\omega^{\mu}\omega_{\mu})
+\frac{1}{2}m_{\rho}^{2}(\rho^{\mu a}\rho_{\mu}^{a})
+\frac{1}{2}m_{\phi}^{2}(\phi^{\mu}\phi_{\mu})
-\frac{1}{4}\Omega^{\mu\nu}\Omega_{\mu\nu}
-\frac{1}{4}\mathbf{R}^{\mu\nu}\mathbf{R}_{\mu\nu}
-\frac{1}{4}\Phi^{\mu\nu}\Phi_{\mu\nu}
+U_{nonl}^{vec}(\omega,\rho,\phi)
+\mathcal{L}_{l},
\end{eqnarray}
%%%%%%%%%%%
% $D_{\mu}=\partial_{\mu}+ig_{B\omega }\omega_{\mu} +ig_{B\phi}\phi_{\mu} +ig_{B\rho}I_{3B}\tau^{a}\rho_{\mu}^{a}$
where the covariant derivative equals $D_{\mu}=\partial_{\mu}+ig_{B\omega }\omega_{\mu} +ig_{B\phi}\phi_{\mu} +ig_{B\rho}\mathbf{I}_{B}\boldsymbol\rho_{\mu}$, $\mathbf{I}_{B}$ denotes isospin of baryon $B$. The baryon effective mass is defined as follows $m_{eff,B}=m_{B} -g_{B\sigma}\sigma -g_{B\sigma^{\ast}}\sigma^{\ast}$ while $\Omega_{\mu\nu}$,  $\mathbf{R}_{\mu\nu}$, and $\Phi_{\mu\nu}$ are the field tensors of the $\omega$, $\rho$, and $\phi$ mesons. A proper description of hyperon--hyperon interaction requires the presence of hidden strangeness mesons: scalar ($\sigma^{\ast}$) and vector ($\phi$).
The Lagrangian function (\ref{lag1}) describes the $\beta$-equilibrated neutron star matter; thus there is also a need to consider the Lagrangian of free leptons  $\mathcal{L}_{l}$
%%%%%%%%%%%
\begin{equation}
\mathcal{L}_{l}=\sum_{l=e,\mu}\overline{\psi}_{l}(i\gamma^{\mu}\partial_{\mu}-m_{l})\psi_{l},
\end{equation}
%%%%%%%%%%%
All nonlinear vector meson couplings that occur in this model have been brought together in the form of a vector potential
%%%%%%%%%%%
\begin{eqnarray}
U_{nonl}^{vec}(\omega, \rho, \phi)
&=&\frac{1}{4}c_{3}(\omega^{\mu}\omega_{\mu})^{2}
+\frac{1}{4}c_{3}(\rho^{\mu a}\rho_{\mu}^{a})^{2}
+\frac{1}{8}c_{3}(\phi^{\mu}\phi_{\mu})^{2}
+\Lambda_{V}(g_{N\omega}g_{N\rho})^{2}(\omega^{\mu}\omega_{\mu})(\rho^{\mu a}\rho^{a}_{\mu})
+\\ \nonumber  
&+&\frac{1}{2}\left(\frac{3}{2}c_{3}-\Lambda_{V}(g_{N\omega}g_{N\rho})^{2}\right)(\phi^{\mu}\phi_{\mu})(\omega^{\mu}\omega_{\mu}+\rho^{\mu a}\rho_{\mu}^{a})
+\frac{1}{4}\Lambda_{V}(g_{N\omega}g_{N\rho})^{2}(\phi^{\mu}\phi_{\mu})^{2}.
\end{eqnarray}
%%%%%%%%%%%
\end{widetext}
%%%%%%%%%%%
The description of dense, hyperon-rich nuclear matter given by the Lagrangian (\ref{lag1})
in the case of non-strange matter is reduced to the standard TM1 model with an extended isovector sector.
This extension refers to the presence of the   $\omega - \rho$  meson coupling and  enables modification of the high density limit of the symmetry energy. The strength of this coupling is characterised by parameter $\Lambda_{V}$. For each value of parameter $\Lambda_{V}$, the parameter $g_{N \rho}$ has to be adjusted to reproduce the symmetry energy $E_{sym} = 25.68$ MeV at $k_{F}=1.15$~fm$^{-1}$~\cite{Horowitz2001}.

%%%%%%%%%%%%%%%%%%%%%%%%%%%%%%%%%%%%%%%%%%%%%%%%%%%%%%%%%%%%%%%%%%%%%%
\section{The equation of state}
%%%%%%%%%%%%%%%%%%%%%%%%%%%%%%%%%%%%%%%%%%%%%%%%%%%%%%%%%%%%%%%%%%%%%%
The mean field approach has been adopted to calculate the EoS. In this approximation, meson fields are separated into classical mean field values: $s_{0}, s_{0}^{\ast}, w_{0}, r_{0}, f_{0}$ and quantum fluctuations, which are neglected in the ground state:
%%%%%%%%%%%
\begin{center}
$\sigma =  \tilde{\sigma}+s_{0}$, \qquad
$\sigma^{\ast}=\tilde{\sigma}^{\ast}+s_{0}^{\ast}$\\
$\omega_{\mu} = \tilde{\omega}_{\mu}+w_{0}$,\qquad 
$w_{0}\equiv <\omega_{\mu}>\delta_{0\mu}=<\omega_{0}>$\\
$\rho_{\mu}^{a} = \tilde{\rho}_{\mu}^{a}+r_{0}$, \qquad  
$r_{0}\equiv <\rho_{\mu}^{a}>\delta_{0\mu}\delta^{3a}=<\rho_{0}>$\\
$\phi_{\mu} = \tilde{\phi}_{\mu}+f_{0}$, \qquad 
$ f_{0}\equiv <\phi_{\mu}>\delta_{0\mu}=<\phi_{0}>$\\
\end{center}
%%%%%%%%%%%
The preferable attribute  of the considered model is its very diverse vector meson sector, which allows one to study the
relevance of different  vector meson couplings for the form of the EoS.
The Lagrangian function (\ref{lag1}) makes it possible to calculate the equations of motion from the corresponding Euler-Lagrange equations.
The obtained results, written in the mean field approximation, take the form:
%%%%%%%%%%%
\begin{widetext}
%%%%%%%%%%%
\begin{equation}
m_{\sigma}^{2}s_{0} =
-g_{3}s_{0}^{2}-g_{4}s_{0}^{3}+\sum_{B}g_{B\sigma
}\frac{2J_{B}+1}{2\pi^2}\int_{0}^{k_{F,B}}\frac{m_{eff,B}(s_{0},s_{0}^{\ast})k^2}{\sqrt{(k^{2}+m_{eff,B}^{2}(s_{0},s_{0}^{\ast}))}}
\label{eqm1}
\end{equation}
%%%%%%%%%%%
\begin{equation}
m_{\sigma^{\ast}}^{2}s_{0}^{\ast} = \sum_{B}g_{B\sigma^{\ast}
}\frac{2J_{B}+1}{2\pi^2}\int_{0}^{k_{F,B}}\frac{m_{eff,B}(s_{0},s_{0}^{\ast})k^2}{\sqrt{(k^{2}+m_{eff,B}^{2}(s_{0},s_{0}^{\ast}))}}
\label{eqm2}
\end{equation}
%%%%%%%%%%%
\begin{equation}
m_{eff,\omega}^{2}w_{0}-2c_{3}w_{0}^{3}=g_{B\omega}\delta_{Q}n_{b},
\label{eqm3}
\end{equation}
%%%%%%%%%%% 	Rozmia
\begin{equation}
m_{eff,\rho}^{2}r_{0}-2c_{3}r_{0}^{3}=g_{B\rho}\delta_{3}n_{b},
\label{eqm4}
\end{equation}
%%%%%%%%%%%
\begin{equation}
m_{eff,\varphi}^{2}f_{0}-(c_{3}+2\Lambda_{V}(g_{B\omega}g_{B\rho})^{2})f_{0}^{3}=g_{B\varphi}\delta_{S}n_{b},
\label{eqm5}
\end{equation}
%%%%%%%%%%%
\begin{equation}
(i\gamma^{\mu}\partial_{\mu}-m_{eff,B}-g_{B\omega}(1-x_{B})\gamma^{0}w_{0}-g_{B \rho}I_{3B}\gamma^{0}\tau^{3}r_{0}-g_{B\phi}x_{B}\gamma^{0}f_{0})\psi_{B}=0,
\end{equation}
%%%%%%%%%%%
where $J_{B}$ and $I_{3B}$ denote the spin and isospin projection of baryon $B$, $m_{eff,i}$,  ($i=\omega, \rho, \phi$)  are effective
masses assigned to vector meson fields:
%%%%%%%%%%%
\begin{equation}
m_{eff,\omega}^{2}=m_{\omega}^{2}+2\Lambda_{V}(g_{B\omega}g_{B\rho})^{2}r_{0}^{2}+\left(\frac{3}{2}c_{3}-\Lambda_{V}(g_{B\omega}g_{B\rho})^{2}\right)f_{0}^{2}+3c_{3}w_{0}^{2}
\end{equation}
%%%%%%%%%%%
\begin{equation}
m_{eff,\rho}^{2}=m_{\rho}^{2}+2\Lambda_{V}(g_{B\omega}g_{B\rho})^{2}w_{0}^{2}+\left(\frac{3}{2}c_{3}-\Lambda_{V}(g_{B\omega}g_{B\rho})^{2}\right)f_{0}^{2}+3c_{3}r_{0}^{2}
\end{equation}
%%%%%%%%%%%
\begin{equation}
m_{eff,\varphi}^{2}=m_{\varphi}^{2}+\left(\frac{3}{2}c_{3}-\Lambda_{V}(g_{B\omega}g_{B\rho})^{2}\right)(w_{0}^{2}+r_{0}^{2})+f_{0}^{2}\left(\frac{3}{4}c_{3}+\Lambda_{V}(g_{B\omega}g_{B\rho})^{2}\right).
\end{equation}
%%%%%%%%%%%
\end{widetext}
%%%%%%%%%%%
Analysis of the equations of motion  raises the issue of the medium effects on the  hadronic matter properties that lead to the  problem of the reduced effective baryon masses $m_{eff,B}$.
Considering the case in which nucleon mass depends only on non-strange condensate and referring to the Walecka model, the relation for the effective baryon masses in the medium in the relativistic mean field description \cite{rmib} can be formulated as
%%%%%%%%%%%
\begin{eqnarray}
\label{eq:meff}
m_{eff,B}&=&m_{eff,B}(s_{0},s_{0}^{\ast})=\\
\nonumber
&=&m_{B}-g_{B\sigma}s_{0}-g_{B\sigma^{\ast}}s_{0}^{\ast}
\end{eqnarray}
%%%%%%%%%%%
where the terms $g_{B\sigma}s_{0}$ and $g_{B \sigma^{\ast}}s_{0}^{\ast}$
represent the modification of baryon masses due to the medium. \\
The source terms in equations (\ref{eqm3}-\ref{eqm5}) can be expressed by introducing the parameters $\delta_{Q}$ and  $\delta_{S}$,
which measure the contributions of strange ($x_{B}$) and non-strange quarks:
%%%%%%%%%%%
\begin{eqnarray}
 &&\delta_{Q}=\sum_{B}(1-x_{B})\frac{n_{B}}{n_{b}}\\
 &&\delta_{3}=\sum_{B}I_{3B}\frac{n_{B}}{n_{b}}\\
 &&\delta_{S}=\sum_{B}x_{B}\frac{n_{B}}{n_{b}}.
\end{eqnarray}
%%%%%%%%%%%
Thus, the number density of $u$ and $d$ quarks refers to $\delta_{Q}$, whereas the number density of strange quarks is given in terms of the parameter $\delta_{S}$.
The parameter $\delta_{3}$ aims to evaluate the difference between the density of $u$ and $d$ quarks, $ n_{b}=\sum_{B}n_{B}$ denotes the total baryon number density.
The baryon--vector meson coupling constants determined from the symmetry relations have been summarised in Table \ref{tab:vector_mesons}.

The numerical solution of the equations of motion, which depends on the form of the effective vector potential $U_{nonl}^{vec}$,  has been limited to the result with  only one single real solution. The demand for the existence of one real solution puts constraints on the value of parameter $\Lambda_{V}$.
Such an analysis in the simpler case of symmetric nuclear matter  leads to an equation that relates $\Lambda_{V}$ and $c_{3}$ parameters and enables the critical value of parameter $\Lambda_{V,cr}$ to be estimated
%%%%%%%%%%%
\begin{equation}
-\frac{7}{4}c_{3}^{2}+4\Lambda_{V}c_{3}(g_{\omega}g_{\rho})^{2}-\Lambda_{V}^{2}(g_{\omega}g_{\rho})^{4}=0.
\end{equation}
%%%%%%%%%%%
The existence of one single real solution is not satisfied for   $\Lambda_{V}>\Lambda_{V,cr}$.
The critical value of the parameter $\Lambda_{V,cr}$ calculated for the selected parameterisations are collected in Table \ref{tab:delta}.
%%%%%%%%%%%
\begin{table}
\caption{Baryon--vector meson coupling constants, $g_{B \omega}=(1-x_{B})g_{N\omega}$ and $x_{B}$ counts the contribution of strange quarks, 
$g_{N\omega} \equiv g_{\omega}$ and $g_{N\rho} \equiv g_{\rho}$.}
\begin{ruledtabular}
\begin{tabular}{ccccc}
Baryon (B) & $x_{B}$ & $g_{B \omega}$  & $g_{B \phi}=x_{B}g_{N\omega}$&$g_{B\rho}$\\
\hline
n  & 0  & $g_{\omega}$ & 0&$g_{\rho}$\\
\hline
p  & 0&  $g_{\omega}$ & 0&$g_{\rho}$\\
\hline
$\Lambda$  & $\frac{1}{3}$  & $\frac{2}{3}g_{\omega}$  & $-\frac{\sqrt{2}}{3}g_{\omega}$&0\\
\hline
$\Sigma$  & $\frac{1}{3}$  & $\frac{2}{3}g_{\omega}$  & $-\frac{\sqrt{2}}{3}g_{\omega}$&$2g_{\rho}$\\
\hline
$\Xi$  & $\frac{2}{3}$  & $\frac{1}{3}g_{\omega}$  &-2$\frac{\sqrt{2}}{3}g_{\omega}$&$g_{\rho}$
\label{tab:vector_mesons}
\end{tabular}
\end{ruledtabular}
\end{table}
%%%%%%%%%%%
\begin{table}
\caption{\label{tab:delta}The critical value of the parameter $\Lambda_{V,cr}$ calculated for the selected parameterisations and the incompressibility $K_{0}$ taken at the saturation density.}
\begin{ruledtabular}
\begin{tabular}{cccc}
Parameter set& $c_{3}$ & $\Lambda_{V,cr}$  & $K_{0}$ (MeV)\\
\hline
NL3 \cite{NL3} & 0  & - & 271\\
\hline
FSUGold \cite{FSUGold}  & 418.39&  0.0517 & 230\\
\hline
TMA \cite{TMA} & 151.59 & 0.0318 & 318\\
\hline
TM1$^{\ast}$ \cite{DelEstal} & 134.624  & 0.0215  & 281.1\\
\hline
TM1 \cite{Sugahara1994} &  71.3 & 0.0156 &281.1\\ \hline
TM2 \cite{TM2} & 84.5318 & 0.0186  &343.8\\
\end{tabular}
\end{ruledtabular}
\end{table}
%%%%%%%%%%%
In order to calculate the energy density and pressure of the nuclear matter, the energy momentum tensor $T_{\mu\nu}$, which is given by the relation
%%%%%%%%%%%
\begin{equation}
T_{\mu\nu}=\frac{\partial{\mathcal{L}}}{\partial({\partial_{\mu}{\varphi_{i}}})}\partial^{\nu}{\varphi_{i}}-\eta_{\mu\nu}\mathcal{L}
\label{tensor}
\end{equation}
%%%%%%%%%%%
has to be used. In equation (\ref{tensor}) $\varphi_{i}$ denotes the boson and fermion fields. The energy density ${\cal E}$ is equal to $<T_{00}>$, whereas the pressure ${\cal P}$ is related to the statistical average of the trace of the spatial component $T_{ij}$ of the energy momentum tensor.
Calculations done for
the considered model lead to the following explicit formulas for
the energy density and pressure:
%%%%%%%%%%%
\begin{widetext}
%%%%%%%%%%%
\begin{eqnarray}
\label{eq:energy}
{\cal E}
&=&\frac{1}{2}m_{\omega}^{2}w_{0}^{2}
+\frac{1}{2}m_{\rho}^{2}r_{0}^{2}
+\frac{1}{2}m_{\phi}^{2}f_{0}^{2}
+\frac{1}{2}m_{\sigma}^{2}s_{0}^{2}
+\frac{1}{3}g_{3}s_{0}^{3}+\frac{1}{4}g_{4}s_{0}^{4}
+\frac{1}{2}m_{\sigma^{\ast}}^{2}s_{0}^{\ast2}
+\frac{3}{4}c_{3}\left(w_{0}^{4}+r_{0}^{4}\right)
+\\ \nonumber 
&+&3\Lambda_{V}(g_{\omega}g_{\rho})^{2}w_{0}^{2}r_{0}^{2}
+\frac{3}{2}\left(\frac{3}{2}c_{3}-\Lambda_{V}(g_{\omega}g_{\rho})^{2}\right)(w_{0}^{2}+r_{0}^{2})f_{0}^{2}
+3\left(\frac{1}{8}c_{3}+\frac{1}{4}\Lambda_{V}(g_{\omega}g_{\rho})^{2}\right)f_{0}^{4}
+\\ \nonumber
&+&\sum_{B}\frac{2}{\pi^{2}}\int_{0}^{k_{F,B}}k^{2}dk\sqrt{k^{2}+m_{eff,B}^{2}}
+{\cal E}_{L},
\end{eqnarray}
%%%%%%%%%%%
\begin{eqnarray}
\label{eq:pressure}
{\cal P}
&=&\frac{1}{2}m_{\omega}^{2}w_{0}^{2}
+\frac{1}{2}m_{\rho}^{2}r_{0}^{2}
+\frac{1}{2}m_{\phi}^{2}f_{0}^{2} 
-\frac{1}{2}m_{\sigma}^{2}s_{0}^{2}
-\frac{1}{3}g_{3}s_{0}^{ 3}
-\frac{1}{4}g_{4}s_{0}^{4}
-\frac{1}{2}m_{\sigma^{\ast}}^{2}s_{0}^{\ast 2}
+\frac{1}{4}c_{3}(w_{0}^{4}+r_{0}^{4})
+\\ \nonumber
&+&\Lambda_{V}(g_{\omega}g_{\rho})^{2}w_{0}^{2}r_{0}^{2}
+\frac{1}{2}\left(\frac{3}{2}c_{3}-\Lambda_{V}(g_{\omega}g_{\rho})^{2}\right)f_{0}^{2}(w_{0}^{2}+r_{0}^{2})
+\left(\frac{1}{8}c_{3}+\frac{1}{4}\Lambda_{V}(g_{\omega}g_{\rho})^{2}\right)f_{0}^{4}
+\\ \nonumber
&+&\sum_{B}\frac{1}{3\pi^{2}}\int_{0}^{k_{F,B}}\frac{k^{4}dk}{\sqrt{k^{2}+m_{eff,B}^{2}}}
+{\cal P}_{L}
\end{eqnarray}
%%%%%%%%%%%
\end{widetext}
%%%%%%%%%%%
where ${\cal E}_{L}$ and ${\cal P}_{L}$ denote the contributions
coming from  leptons.
%%%%%%%%%%%%%%%%%%%%%%%%%%%%%%%%%%%%%%%%%%%%%%%%%%%%%%%%%%%%%%%%%%%%%%
\section{Coupling constants}
%%%%%%%%%%%%%%%%%%%%%%%%%%%%%%%%%%%%%%%%%%%%%%%%%%%%%%%%%%%%%%%%%%%%%%
\subsection{Strange baryons}
%%%%%%%%%%%%%%%%%%%%%%%%%%%%%%%%%%%%%%%%%%%%%%%%%%%%%%%%%%%%%%%%%%%%%%
Understanding the nature of interactions between baryons is a decisive factor for the properties of neutron stars.
A precise description of baryon interactions in the strange sector of the model is essential for the correct construction of the EoS.
However, the incompleteness of the experimental data intensifies the uncertainties that are connected with the evaluation of coupling constants that involve strange baryons. 
In general, there are very few data available to describe hyperon--nucleon (YN) and hyperon--hyperon (YY) interactions.
Hyperon--vector meson coupling constants are taken from the quark model. They are summarised in Table \ref{tab:vector_mesons}. In the scalar sector, the scalar couplings $g_{B\sigma}$ of the $\Lambda$, $\Sigma$ and $\Xi$ hyperons require constraining in order to reproduce the estimated values of the potentials felt by a single $\Lambda$, $\Sigma$ and $\Xi$ in the saturated nuclear matter.
In the case of $\Lambda$ hypernuclei there is a considerable amount of data on binding energies and single particle levels allowing the identification of the potential felt by a single $\Lambda$ in nuclear matter in the range $U^{(N)}_{\Lambda} \approx 27-30$ MeV \cite{Hasegawa1996,Hotchi2001}. 
Considering the $\Xi$ hypernuclei current knowledge about the interaction of $\Xi$ hyperons with nuclei is very limited.
Dover and Gal \cite{Dover1983} based on early emulsion data indicated an attractive $\Xi$-nucleus potential of the order of 21-24 MeV. 
This result agrees with theoretical predictions for $\Xi$ in nuclear matter obtained in the model D of the Nijmegen group \cite{Nagels1975}.
However, the missing-mass spectra of a double-charge exchange reaction $(K^{-},K^{+})$ on a ${}^{12}C$ target have suggested the $\Xi$ well depth of 14-16 MeV \cite{Khaustov2000,Fukuda1998}.
Analysis of the experimental data indicate a repulsive  $\Sigma$-nucleus potential \cite{Noumi2002}, with a substantial isospin dependence \cite{Nagae1998}.
In the case of YY interactions, the only sources of information are the double-strange hypernuclear systems.
Several events have been identified that suggest an attractive $\Lambda\Lambda$ interaction.
The most promising results, known as the NAGARA event \cite{Takahashi2001} with the ${}_{\Lambda\Lambda}^{6}$He hypernucleus, indicate that the
$\Lambda\Lambda$ interaction is weakly attractive. The estimated value of the $U_{\Lambda}^{(\Lambda)}$ potential at the level of 5 MeV permits a parameter set which reproduces this weaker $\Lambda\Lambda$ interaction to be estimated \cite{hypnonlin}.

The potential that describe hyperon--nucleon and hyperon--hyperon interaction can be written in the form that involves both the scalar and
vector coupling constants 
%%%%%%%%%%%
\begin{eqnarray}
U_{Y}^{(B)}
=g_{Y\sigma}s_{0}
-g_{Y\omega}w_{0}
+g_{Y\sigma^{\ast}}s_{0}^{\ast}
-g_{Y\phi}f_{0}
=\nonumber \\
=m_{Y}
-m_{eff,Y}(s_{0},s_{0}^{\ast})
-(g_{Y\omega}w_{0}
+g_{Y\phi}f_{0}),
\end{eqnarray}
%%%%%%%%%%%
where $m_{eff,Y}(s_{0},s_{0}^{\ast})$ is the effective mass and  $Y$ stands for the $\Lambda$,
$\Sigma$ and $\Xi$ hyperons. For the determination of the
$g_{\Lambda\sigma}$, $g_{\Sigma \sigma}$ and $g_{\Xi\sigma}$
coupling constants, the following values of the potentials were
used  
%%%%%%%%%%%
\begin{equation} 
U_{\Lambda}^{(N)}=-28 \text{MeV},\,\,
U_{\Sigma}^{(N)}+30 \text{MeV},\,\, 
U_{\Xi}^{(N)}=-18 \text{MeV}.
\end{equation}
%%%%%%%%%%%
In general the coupling constants $g_{Y\sigma}$  can be decomposed into two parts $g_{Y\sigma}=g_{\sigma}x_{Y\sigma}$, where $x_{Y\sigma}$ depends on the value of potential $U^{(N)}_{Y}$.
 The coupling of hyperons to the strange meson $\sigma^{\ast}$ \cite{Schaffner1995} were
obtained from the following relations 
%%%%%%%%%%%
\begin{equation}
U_{\Xi}^{(\Xi)}\simeq U_{\Lambda}^{(\Xi)}\simeq2U_{\Xi}^{(\Lambda)}\simeq2U_{\Lambda}^{(\Lambda)}.
\end{equation}
%%%%%%%%%%%
The scalar coupling constants are collected in Tables \ref{tab:scalar} and \ref{tab:sigma}.
%%%%%%%%%%%
\begin{table}
\caption{The scalar coupling constants.}
\label{tab:scalar}
\begin{ruledtabular}
\begin{tabular}{cccccc} 
$g_{\Lambda \sigma}$ & $g_{\Xi \sigma}$ & $g_{\Sigma \sigma}$ & $g_{\Lambda \sigma^{\ast}}$ & $g_{\Xi \sigma^{\ast}}$ & $g_{\Sigma \sigma^{\ast}}$ \\ 
  6.169 & 3.201 & 4.476 & 5.262 & 11.623 & 5.626 \\ 
\end{tabular} 
\end{ruledtabular}
\end{table}
%%%%%%%%%%%
\begin{table}
\caption{Scalar coupling constants calculated for chosen values of the potentials $U^{(N)}_{\Sigma}$ and $U^{(N)}_{\Xi}$.}
\label{tab:sigma}
\begin{ruledtabular}
\begin{tabular}{ccccccc} 
$U^{(N)}_{\Sigma}$ & $U^{(N)}_{\Xi}$ & $g_{\Lambda \sigma}$ & $g_{\Xi \sigma}$ & $g_{\Sigma \sigma}$  & $g_{\Xi \sigma^{\ast}}$ & $g_{\Sigma \sigma^{\ast}}$ \\
(MeV) & (MeV) &  &  &  &  &  \\ \hline \hline
+30 & -14 & 6.169 & 3.084 & 4.476 & 5.626 & 11.474 \\ \hline
+30 & -18 & 6.169 & 3.201 & 4.476 & 5.482 & 11.372 \\ \hline
+20 & -18 & 6.169 & 3.201 & 4.768 & 5.482 & 11.372 \\ \hline
+10 & -18 & 6.169 & 3.201 & 5.060 & 5.482 & 11.372 \\ \hline
-10 & -18 & 6.169 & 3.201 & 5.644 & 5.482 & 11.372 \\ \hline
-20 & -18 & 6.169 & 3.201 & 5.935 & 5.482 & 11.372 \\ \hline
-30 & -18 & 6.169 & 3.201 & 6.227 & 5.482 & 11.372 \\ 
\end{tabular} 
\end{ruledtabular}
\end{table}
%%%%%%%%%%%
%%%%%%%%%%%%%%%%%%%%%%%%%%%%%%%%%%%%%%%%%%%%%%%%%%%%%%%%%%%%%%%%%%%%%%
\subsection{The symmetry energy}
%%%%%%%%%%%%%%%%%%%%%%%%%%%%%%%%%%%%%%%%%%%%%%%%%%%%%%%%%%%%%%%%%%%%%%
The TM1 parameter set \cite{Sugahara1994} that successfully describes the ground state properties of both finite nuclei and  infinite nuclear matter was  supplemented with the mixed nonlinear isoscalar--isovector $\omega - \rho$ meson coupling,  which provides the additional possibility of modifying the high density components of the symmetry energy. The remaining nuclear matter ground state properties  were left unchanged.
The symmetry energy is given by the relation
%%%%%%%%%%%
\begin{equation}
E_{sym}(n_b)=\frac{k_{F}^{2}}{6\sqrt{k_{F}^{2}+m_{eff,N}^{2}}}+\frac{g_{\rho}^2}{12\pi^{2}}\frac{k_{F}^{3}}{\hat{m}^{2}_{eff,\rho}},
\end{equation}
%%%%%%%%%%%
where $\hat{m}^{2}_{eff,\rho}=m^{2}_{eff,\rho}(f_{3}=0)$ denotes the effective mass of the $\rho$ meson in the case of non-strange, symmetric nuclear matter.

The strength of the $\omega - \rho$ coupling is set by the $\Lambda_{V}(g_{\omega}g_{\rho})^{2}$ and requires the adjustment of the $g_{\rho}$
coupling constant to keep the empirical value  of the symmetry energy $E_{sym}(n_{b})=25.68$~MeV at the baryon density $n_{b}$, which corresponds to $k_{F}=1.15$~fm$^{-1}$ \cite{Horowitz2001}.
The  parameters $\Lambda_{V}$ together with the adjusted value of the parameter $g_{\rho}$ are presented in Table \ref{tab:tm1set}.
%%%%%%%%%%%
\begin{table}[h]
\caption{TM1 parameter set with the extended isovector sector \cite{Sugahara1994}. }
\label{tab:tm1set}
\begin{ruledtabular}
\begin{tabular}{lll} 
\multicolumn{3}{c}{TM1} \\ \hline
&& \\
$m_{\sigma}$ = 511.2 MeV  & $g_{\sigma}$ = 10.029 & $g_{3}$ = 7.2325~fm$^{-1}$ \\ %\hline
$m_{\omega}$ = 783 MeV & $g_{\omega}$ = 12.614 & $g_{4}$ = 0.6183 \\ %\hline
$m_{\rho}$ = 770 MeV & $g_{\rho}$ = 9.264 & $c_{3}$ = 71.0375  
\end{tabular} 
\begin{tabular}{ccccccc} 
\multicolumn{7}{c}{TM1 nonlinear (isovector sector)}  \\  \hline
&&&& \\  
$\Lambda_{V}$ & 0 & 0.014 & 0.015 & 0.016 & 0.0165 & 0.017 \\ %\hline
$g_{\rho}$ & 9.264 & 9.872 & 9.937 & 10.003 & 10.037 & 10.071  \\ 
&&&&&& \\ 
$L$ (MeV) & 108.58 & 77.52 & 75.81 & 74.16 & 73.36 & 72.56 \\ 
\end{tabular} 
\end{ruledtabular}
\end{table}
%%%%%%
The density dependence of the symmetry energy can be expressed by coefficients that define the slope $L$ and the curvature $K_{sym}$ of the symmetry energy
\begin{equation}
\label{eq:slope}
E_{sym}(n_b)=E_{sym}(n_0)+L\left(\frac{n_{b}-n_{0}}{3n_{0}}\right)+\frac{K_{sym}}{2!}\left(\frac{n_{b}-n_{0}}{3n_{0}}\right)^2.
\end{equation}
Equation (\ref{eq:slope}) is a typical low density expansion,
higher-order terms must be taken into account at suprasaturation densities.
The performed calculations that were focused mainly on the slope parameter of the symmetry energy 
\begin{equation}
L=3n_{0}\frac{\partial^{2}E_{sym}(n_b)}{\partial{n}_{b}^{2}}|_{n_{b}=n_{0}}
\end{equation}
gave results that agreed with the experimental data ($L=88\pm 25$~MeV \cite{Agrawal2010}). These results are collected in Table~\ref{tab:tm1set}. However, current estimation of the symmetry energy parameter based on theoretical, experimental and observational results narrows the range of $L$ to $(40.5-61.9)$~MeV \cite{Lattimer2013}.
%%%%%%%%%%%
%%%%%%%%%%%%%%%%%%%%%%%%%%%%%%%%%%%%%%%%%%%%%%%%%%%%%%%%%%%%%%%%%%%%%%
\section{Results}
\begin{figure}
\includegraphics[clip,width=8.57cm]{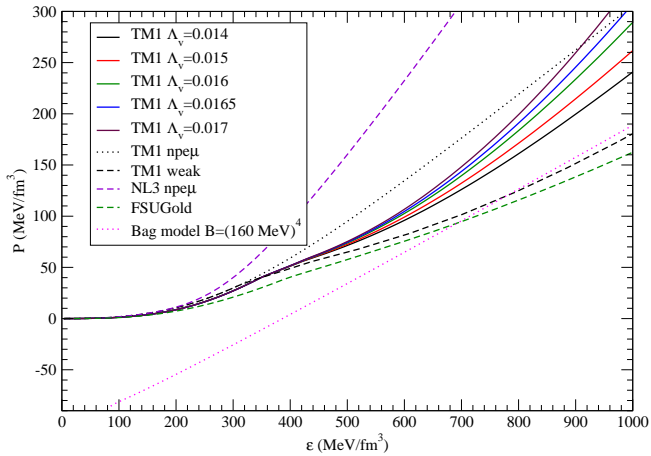} % eosTM1.eps: 1048592x1048592 pixel, 300dpi, 8878.08x8878.08 cm, bb=(atend)
\caption{(Color online) The pressure vs density calculated for selected  parameterisations. The stiffest and softest EoS have been obtained for the non-strange
matter for NL3 and FSUGold. The EoSs calculated for TM1 parameter set form a distinct class. For these EoSs the stiffest  is the one for non-strange matter. It is represented by the dotted line. The nonlinear models for different values of the coupling strength $\Lambda_{V}$ are also presented. The upper limit for these models  is for $\Lambda_{V}=0.017$  the softest represents the case of the standard TM1
model extended by the inclusion of strange mesons which have been introduced in a minimal fashion. The remaining EOS for the nonlinear
models with different values of parameter $\Lambda_{V}$ are located between these two curves.}
\label{fig:eos}
\end{figure}

%%%%%%%%%%%%%%%%%%%%%%%%%%%%%%%%%%%%%%%%%%%%%%%%%%%%%%%%%%%%%%%%%%%%%%
The energy density and pressure given by relations (\ref{eq:energy}) and  (\ref{eq:pressure})
define the EoS. Numerical calculations, which were made for the TM1 parameterisation for both nonstrange and strangeness-rich matter, led to the solutions that are shown in Fig.~\ref{fig:eos}. A class of EoSs was obtained. Individual EoSs are parametrized by the coupling constant $\Lambda_{V}$, which determines the strength of the mixed vector meson interactions. The form of the EoSs allows one to compare the differences between various models.
%%%%%%%%%%%

%%%%%%%%%%%%%%%%%%%%%%%%
An analysis performed for the nonlinear model with parameter $\Lambda_{V}$, which ranges between 0.014 to 0.017, showed that the increase of $\Lambda_{V}$ produces a stiffer EoS.
For comparison, the EoSs for non-strange matter for NL3, TM1 and FSUGold parameterisations have been included.
Thus, the stiffness of the EoS depends on the existence and strength of the mixed vector meson interactions, and
the nonlinear model described by the Lagrangian function (\ref{lag1}) makes it possible to construct a much stiffer EoS than the one obtained for the TM1-weak model. The  abbreviation TM1-weak denotes the standard TM1 model extended to the full octet of baryons  with two additional meson fields, which were introduced in a minimal fashion, to reproduce the hyperon--hyperon interaction.
In order to make the analysis more complete, the conditions under which the quark matter  occurs in neutron star interiors  was established.  This permits the formation of a stable hybrid star configuration. Two phases of matter were compared: the strange hadronic matter and the quark matter. The phase with the highest pressure (lowest free energy) was favoured. The dotted curve shows the quark matter EoS for the fixed value of the bag parameter $B^{1/4}=160$~MeV.
The result indicates that there is no intersection of the quark matter EoS and that of hyperon-rich matter calculated for the nonlinear model for a different values of parameter $\Lambda_{V}$, and in this case a hybrid star with a quark phase inside cannot be constructed. Solutions that permit the existence of the quark matter phase inside the hyperon star were obtained for the TM1-weak and FSUGold parameterisations.

\begin{figure}
\includegraphics[clip,width=8.67cm]{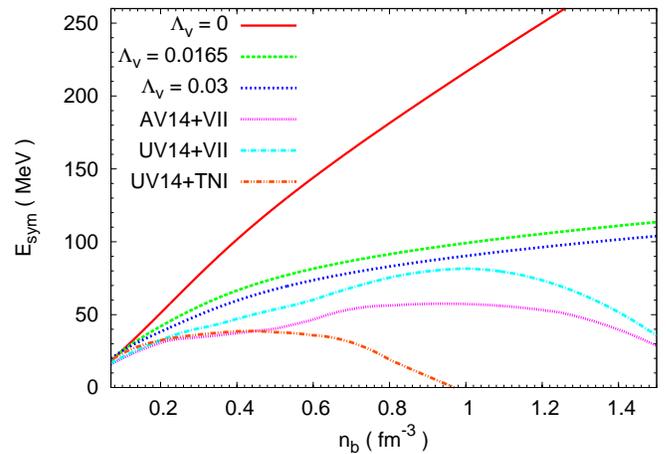} % eosTM1.eps: 1048592x1048592 pixel, 300dpi, 8878.08x8878.08 cm, bb=(atend)
\caption{(Color online) The density dependence of symmetry energy calculated for TM1 parameter set \cite{Sugahara1994} for different values of parameter $\Lambda_{V}$. For comparison the results obtained for the AV14+VII, UV14+VII and UV14+TNI models \cite{Wiringa1988} are included.}
\label{fig:ensym}
\end{figure}

In the case of nuclear matter an extended isovector sector comprises the $\omega - \rho$ meson interaction. Parameter $\Lambda_{V}$ sets the strength of the $\omega - \rho$ coupling.
This term altered the density dependence of the symmetry energy. The standard TM1 parameterisation without $\omega - \rho$ coupling gives as a result very stiff form of the symmetry energy. The inclusion of the $\omega - \rho$ coupling softens the symmetry energy. The solutions were presented in Fig.~\ref{fig:ensym}.  Calculations were done for rather high value of $\Lambda_{V} = 0.0165$ and 0.03. The interaction between $\omega$ and $\rho$ mesons leads to the solution, which approaches that obtained for the AV14 and UV14 models with the Urbana VII (UV VII) three nucleon potential. For comparison the form of the symmetry energy calculated for the UV14 plus TNI model was included \cite{Wiringa1988}.
%%%%%%%%%%%
%%%%%%%%%%%

\begin{figure}
\includegraphics[clip,width=8.57cm]{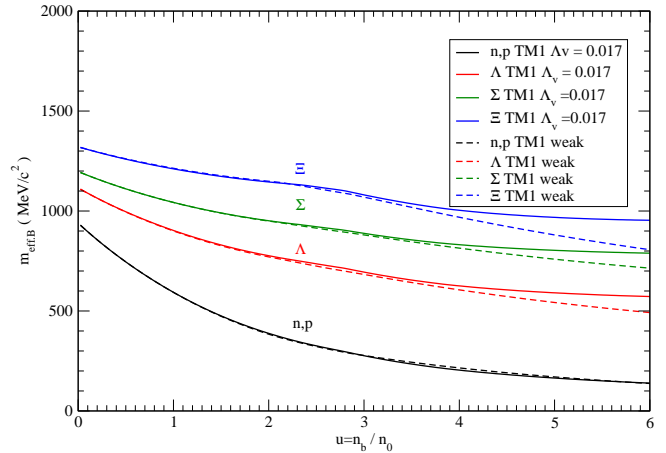} % eosTM1.eps: 1048592x1048592 pixel, 300dpi, 8878.08x8878.08 cm, bb=(atend)
\caption{(Color online) The effective baryon masses as a function of relative baryon  density calculated for the nonlinear model, for $\Lambda_{V} = 0.017$ and for the TM1-weak model.}
\label{fig:meff}
\end{figure}

\begin{figure}
\centering \includegraphics[clip,width=8.57cm]{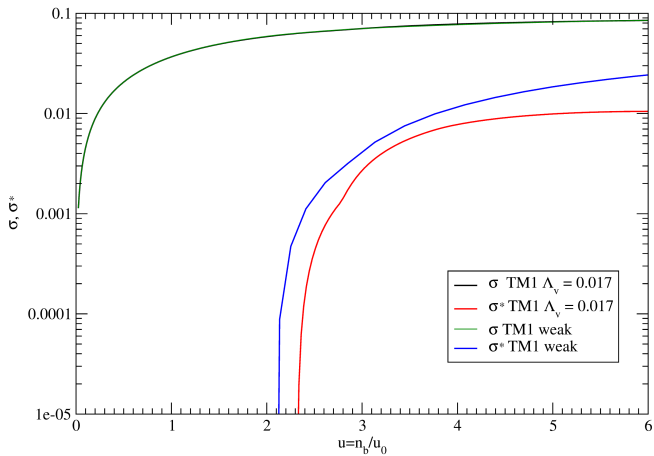} % eosTM1.eps: 1048592x1048592 pixel, 300dpi, 8878.08x8878.08 cm, bb=(atend)
\caption{(Color online) The scalar mesons $\sigma$ and $\sigma^{\ast}$ as a function of baryon density. For comparison the results obtained for the TM1-weak model have been included.}
\label{fig:scalarmesons}
\end{figure}

\begin{figure}
\includegraphics[clip,width=8.57cm]{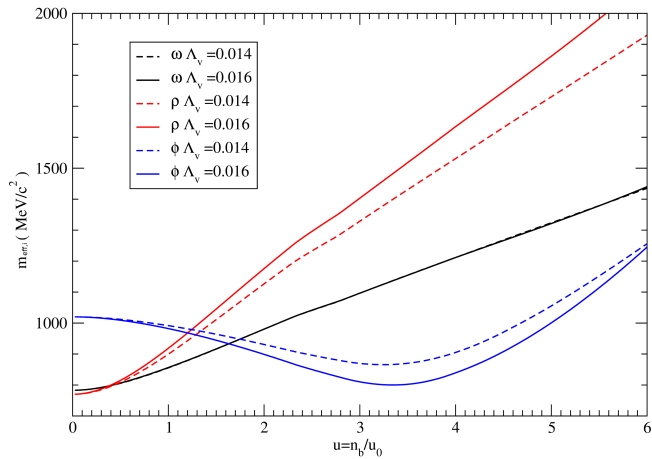} % eosTM1.eps: 1048592x1048592 pixel, 300dpi, 8878.08x8878.08 cm, bb=(atend)
\caption{(Color online) The effective meson masses as a function of baryon number density calculated for the nonlinear model, for $\Lambda_{V} = 0.014$ and $\Lambda_{V} = 0.016$.}
\label{fig:mvector}
\end{figure}

\begin{figure}
\includegraphics[clip,width=8.57cm]{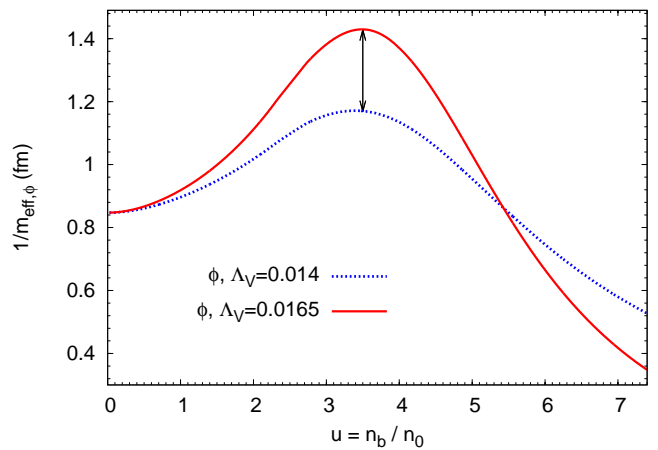} % eosTM1.eps: 1048592x1048592 pixel, 300dpi, 8878.08x8878.08 cm, bb=(atend)
\caption{(Color online) The density dependence of the factor  $1/m_{eff,\phi}$  calculated for the nonlinear model, for $\Lambda_{V} = 0.014$ and $\Lambda_{V} = 0.0165$.}
\label{fig:phi}
\end{figure}

Interacting baryons are the basic components of the matter of neutron stars.
The modification of baryon masses that arises from baryon interactions with the background nuclear matter is shown in Fig.~\ref{fig:meff}.
The numerical solutions predicted by equation (\ref{eq:meff}) for the fixed value of parameter $\Lambda_{V}= 0.017$ for both the nonlinear model and for the TM1-weak model, show reduced effective baryon masses.
%%%%%%%%%%%
%%%%%%%%%%%
The reduction of the nucleon mass in the  nonlinear model is essentially the same as that obtained in the TM1-weak one.
Thus, the influence of the nonlinear vector meson couplings on the nucleon effective mass is negligible.
The effective masses of strange baryons in the case of the nonlinear model drop less rapidly  than the effective masses  obtained in the TM1-weak model. The behaviour of the baryon effective masses is governed by the  density dependence  of the scalar mean fields,
which is presented in Fig.~\ref{fig:scalarmesons}.
The presence of nonlinear couplings between vector mesons modifies the density dependence of the strange scalar meson leaving the nonstrange scalar meson almost unchanged.
%%%%%%%%%%%
%%%%%%%%%%%
The in-medium reduction of baryon masses  is equivalent to the modification of vector meson masses in the meson sector (Fig.~\ref{fig:mvector}). An analysis of the density dependence of the effective $\rho$ and $\phi$ vector meson masses led to the conclusion that their modification was produced by a strong $\Lambda_{V}$ dependence, especially in the high density limit. The effective mass of the $\omega$ meson is almost independent of the value of parameter $\Lambda_{V}$.
%%%%%%%%%%%
%%%%%%%%%%%
The stiffness of the EoS is characterised by the incompressibility of nuclear matter. In general, incompressibility comprises terms resulting from the kinetic pressure of Fermi gas and from the potential of the model. Analysing the strange sector of the model one can compare the factor that determines the difference between the strength of the effective repulsive and attractive forces.
The scalar meson $\sigma^{\ast}$ has been introduced in a minimal fashion thus 
$m_{eff,\sigma^{\ast}}=m_{\sigma^{\ast}}$. The strength of the effective repulsive force between strange baryons is mainly altered by the factor $1/m_{eff,\phi}$. The influence of the $\Lambda_{V}$ parameter on the density dependence of $1/m_{eff,\phi}$ is depicted in  Fig.~\ref{fig:phi}. The increase of the parameter $\Lambda_{V}$ considerably  enhances the strength of the repulsive force in the system.  
%%%%%%%%%%%
%%%%%%%%%%%

\begin{figure}
\includegraphics[clip,width=8.57cm]{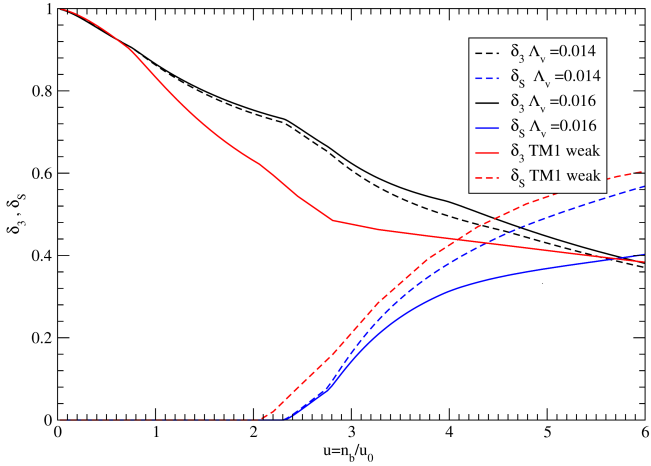} % eosTM1.eps: 1048592x1048592 pixel, 300dpi, 8878.08x8878.08 cm, bb=(atend)
\caption{(Color online) The asymmetry parameters $f_3=\delta_3$ and $f_S=3\delta_S$ as a function of baryon number density calculated for the nonlinear model, for different values of $\Lambda_{V}$ and for the TM1-weak model.}
\label{fig:asym}
\end{figure}

The properties of asymmetric strangeness-rich matter of neutron stars  are characterized by the parameters that define the strangeness content of the system $f_{S}=n_{S}/n_{b}$, where $n_{S}$ is the strangeness density and the isospin asymmetry  $f_{3}=n_{3}/n_{b}$, where $n_{3}$ is the isospin density given by the relation
%%%%%%%%%%%
\begin{equation}
n_{3}=\sum_{B}I_{3 B}n_{B}
\end{equation}
%%%%%%%%%%%
The density dependence of the asymmetry parameter and the strangeness content of the system is depicted in Fig.~\ref{fig:asym}. The nonlinear model leads to a system with an enhanced asymmetry and a considerably reduced strangeness. An increase of  parameter $\Lambda_{V}$ causes  the matter to become more asymmetric.
%%%%%%%%%%%
%%%%%%%%%%%

\begin{figure}
\includegraphics[clip,width=8.67cm]{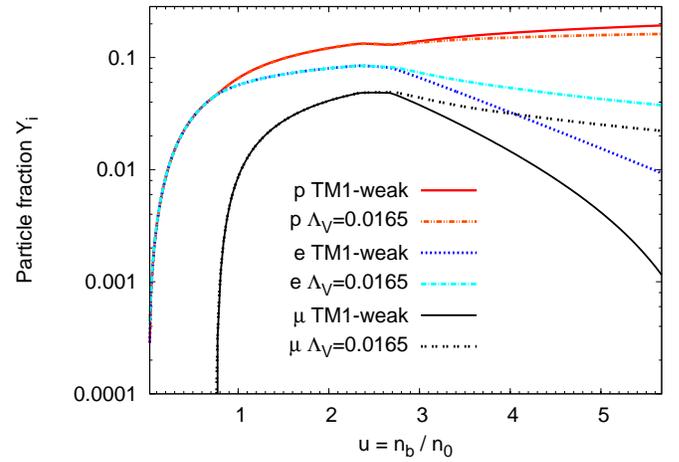} % eosTM1.eps: 1048592x1048592 pixel, 300dpi, 8878.08x8878.08 cm, bb=(atend)
\caption{(Color online) Relative concentrations of nucleons and leptons calculated for the nonlinear model, for the fixed value of parameter $\Lambda_{V}=0.0165$. For comparison the results obtained for the TM1-weak model has been included.}
\label{fig:nonstrange}
\end{figure}

\begin{figure}
\includegraphics[clip,width=8.67cm]{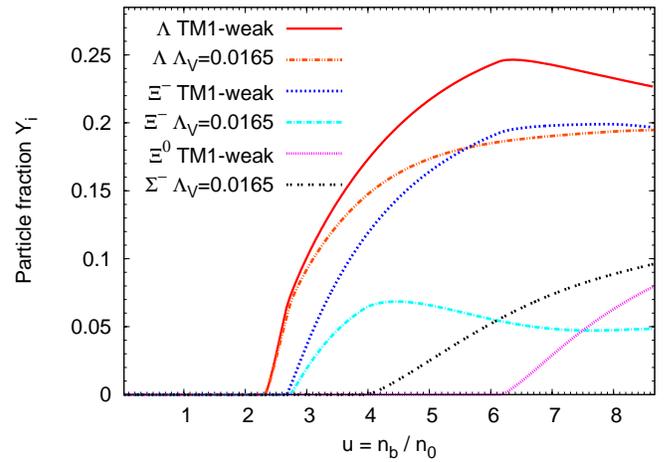} % eosTM1.eps: 1048592x1048592 pixel, 300dpi, 8878.08x8878.08 cm, bb=(atend)
\caption{(Color online) Relative concentrations of strange baryons calculated for the nonlinear model, for the fixed value of parameter $\Lambda_{V}=0.0165$. For comparison the results obtained for the TM1-weak model has been included.}
\label{fig:strange}
\end{figure}

\begin{figure}
\includegraphics[clip,width=8.67cm]{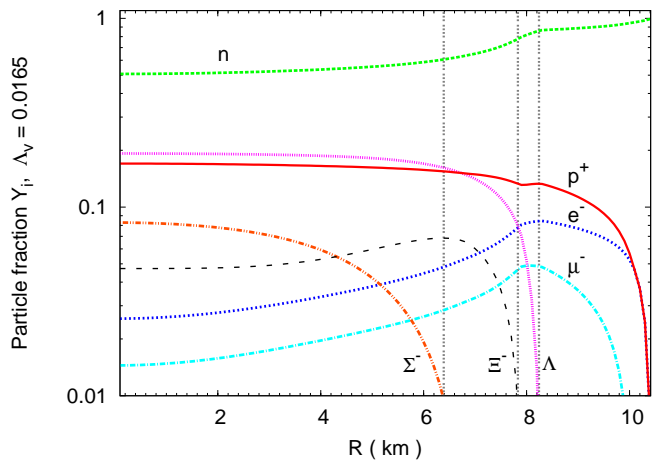} % eosTM1.eps: 1048592x1048592 pixel, 300dpi, 8878.08x8878.08 cm, bb=(atend)
\caption{(Color online) The particle fraction $Y_i$ as a function of the radius of the neutron star, for the maximal mass configuration. Calculations has been done on the basis of nonlinear model with $\Lambda_{V} = 0.0165$. The dotted vertical lines point the threshold densities for particular hyperons.}
\label{fig:particle}
\end{figure}

Charge neutrality and the condition of $\beta$ equilibrium (~$p+e^{-} \leftrightarrow n+\nu_{e}$~) impose constraints on a neutron star composition. Assuming that neutrinos are not considered, since their mean free path is longer then the star radius, the following relation can be obtained
%%%%%%%%%%%
\begin{equation}
\mu_{p}+\mu_{e}=\mu_{n}.
\end{equation}
%%%%%%%%%%%
Additional hadronic states are produced in neutron star interiors at sufficiently high densities when hyperon in-medium energy equals their chemical potential. The higher the density the more various hadronic species are expected to populate. 
In general weak reactions for baryons and the corresponding equations  for chemical potentials can be written in the form
%%%%%%%%%%%
\begin{eqnarray}
\label{eq:hyperons}
&&B_{1}+l\leftrightarrow B_{2}+\nu_{l} \\ \nonumber
&& \mu_{B}=b_{B}\mu_{n}-q_{B}\mu_{e}
\end{eqnarray}
%%%%%%%%%%%
where $B_{1}$ and $B_{2}$ denote baryons, $l$ and $\nu_{l}$ lepton and neutrino of the same flavor, whereas $b_{B}$ and $q_{B}$ refer to baryon $B$ with baryon number $b_{B}$ and charge $q_{B}$.
The above equations create relations between chemical potentials of particular hyperons
%%%%%%%%%%%
\begin{eqnarray}\nonumber
&&\mu_{\Lambda}=\mu_{\Sigma^{0}}=\mu_{\Xi^{0}}=\mu_{n}; \\ 
&&\mu_{\Sigma^{-}}=\mu_{\Xi^{-}}=\mu_{n}+\mu_{e}; \\ \nonumber
&&\mu_{p}=\mu_{\Sigma^{+}}=\mu_{n}-\mu_{e}.
\end{eqnarray}
%%%%%%%%%%%
The presented equilibrium conditions  determine  all constituents of the matter of neutron stars.
The concentrations of a particular component $i=B,l$ of the matter of a neutron star can be defined as $Y_{i}=n_{i}/n_{b}$,  where  $n_{i}$ denotes the density of the component $i$ and $n_{b}$ is the total  baryon number density.
The density fraction of nucleons and leptons as a function of the baryon number density for the fixed value of  parameter $\Lambda_{V}$ is presented in Fig.~\ref{fig:nonstrange}. The important findings concern the concentration of leptons, which are more highly populated in the case of the nonlinear model. 
Fig.~\ref{fig:strange} shows how the modification of the vector meson sector alters concentrations of strange baryons.
The first hyperon that appears is $\Lambda$ and it is followed by $\Xi^{-}$ and $\Sigma^{-}$. The appearance of negatively charged   hyperons reduce the concentrations of leptons. This stems from the charge neutrality condition. However, the initial rapid increase in population of $\Xi^{-}$ hyperons has been suppressed leading to significantly reduced concentration of $\Xi^{-}$ hyperons at sufficiently high density. 
For comparison the results obtained for the TM1-weak model have been included. It is evident that the additional nonlinear couplings between vector mesons modify chemical composition of the neutron star shifting the onset point to higher densities and reduces the strangeness content of the system.
%%%%%%%%%%%
%%%%%%%%%%%
%%%%%%%%%%%
%%%%%%%%%%%
A composition and concentrations of hyperons calculated in the nonlinear model can be traced in a chosen configuration of a neutron star. The composition of the core of the maximum mass configuration is depicted in  Fig.~\ref{fig:particle}. The number density of $\Xi^{-}$ hyperons is reduced. However, an interesting feature of this model is the abundance of $\Sigma^{-}$ hyperons in the very inner part of the neutron star inner core.
%%%%%%%%%%%
%%%%%%%%%%%

The most essential feature of the model is connected with the fact that even in the presence of hyperons the obtained EoS is very stiff.
Global neutron star parameters such as the mass and radius and the structure of a neutron star can be determined by the equation of hydrostatic equilibrium - the Tolman-Oppenheimer-Volkoff (TOV) equation:
%%%%%%%%%%%
\begin{eqnarray}\label{TOV}
\frac{d{\cal P}(r)}{dr}&=&\frac{-G\left({\cal E} (r) + {\cal P}(r)\right)\left(m(r)+4\pi r^3{\cal P}(r)\right)}{r^{2}\left(1-\frac{2Gm(r)}{r}\right)}\\ \nonumber
\frac{dm(r)}{dr}&=&4\pi r^{2}{\cal E} (r) \\ \nonumber
\frac{dn(r)}{dr}&=&4\pi r^{2}\left(1-\frac{2Gm(r)}{r}\right)^{-1/2}
\end{eqnarray}
%%%%%%%%%%%
where $m$ and $n$ denote the enclosed gravitational mass and baryon number, respectively, $\cal{P}$ is the pressure and ${\cal E}$ is the total energy density.
In order to get the numerical solution of equation (\ref{TOV}),  the   EoS has to be specified. A correct model of a neutron star is based on the assumption that its internal structure is composed of separate parts, thus the construction of the mass--radius relation requires taking into account additional EoSs that describe the matter of the inner and outer crust. For the outer and inner crusts the EoSs of Baym, Pethick, and Sutherland (BPS) \cite{BPS} and Baym, Bethe and Pethick (BBP) \cite{BBP} have been used respectively.
%%%%%%%%%%%
\begin{figure}
\includegraphics[clip,width=8.57cm]{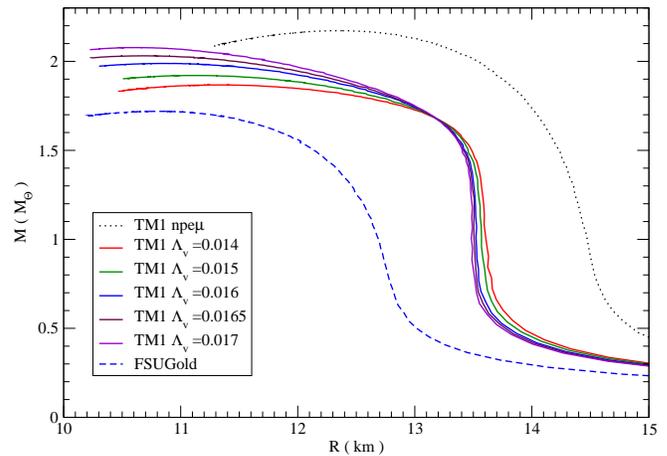} % eosTM1.eps: 1048592x1048592 pixel, 300dpi, 8878.08x8878.08 cm, bb=(atend)
\caption{(Color online) The mass--radius relations calculated for the nonlinear model for different values of parameter $\Lambda_{V}$. The results obtained for both the TM1 parameterisation in the case when the matter of the neutron star comprises only nucleons and for FSUGold parameter set.}
\label{fig:rm}
\end{figure}
%%%%%%%%%%%
The results obtained for the set of EoSs calculated in this paper led to the mass-radius relations and permitted the value of the maximum mass to be determined which in a sense can give a measure of the impact of particular nonlinear couplings between vector mesons. In the mass-radius relations obtained for varying values of  parameter $\Lambda_{V}$ are shown in Fig. \ref{fig:rm}. The higher the value of $\Lambda_{V}$, the higher the maximum mass.\\

%%%%%%%%%%%%%%%%%%%%%%%%%%%%%%%%%%%%%%%%%%%%%%%%%%%%%%%%%%%%%%%%%%%%%%
\section{The influence of the $U^{(N)}_{Y}$ potentials}
\begin{figure}
\includegraphics[clip,width=8.57cm]{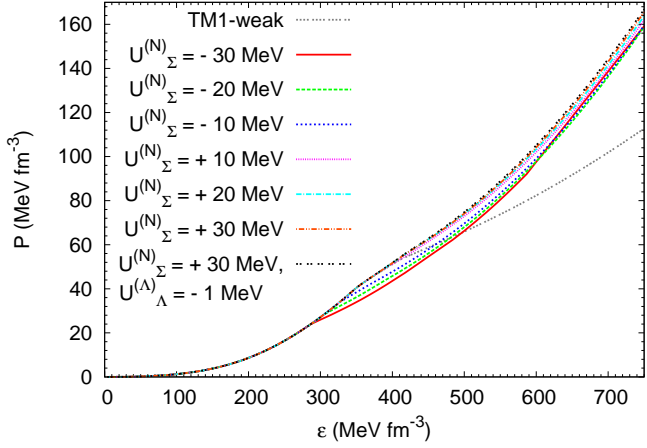} % eosTM1.eps: 1048592x1048592 pixel, 300dpi, 8878.08x8878.08 cm, bb=(atend)
\caption{(Color online) The pressure vs density calculated for different values of the $ U^{(N)}_{\Sigma} $ potential (attractive and repulsive), the case of a very weak $\Lambda - \Lambda$ interaction ($U^{(\Lambda)}_{\Lambda} = 1$ MeV) has been also included. For comparison the EoS calculated for the model TM1-weak has been shown.}
\label{fig:pot}
\end{figure}

\begin{figure}
\includegraphics[clip,width=8.57cm]{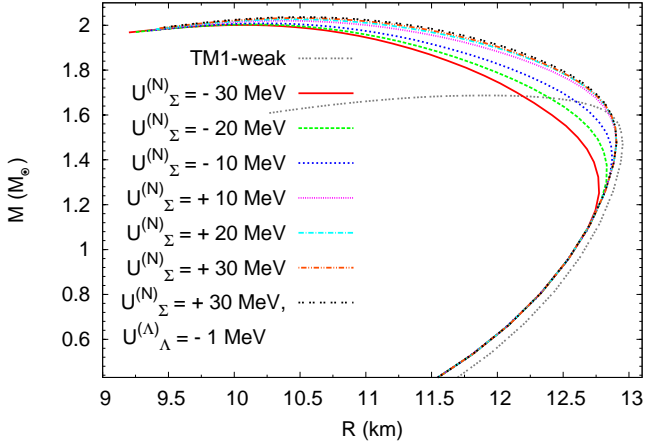} % eosTM1.eps: 1048592x1048592 pixel, 300dpi, 8878.08x8878.08 cm, bb=(atend)
\caption{(Color online) The mass -- radius relations calculated for EoSs presented in Fig.\ref{fig:pot}.}
\label{fig:rm_pot}
\end{figure}

\begin{figure}
\includegraphics[clip,width=8.57cm]{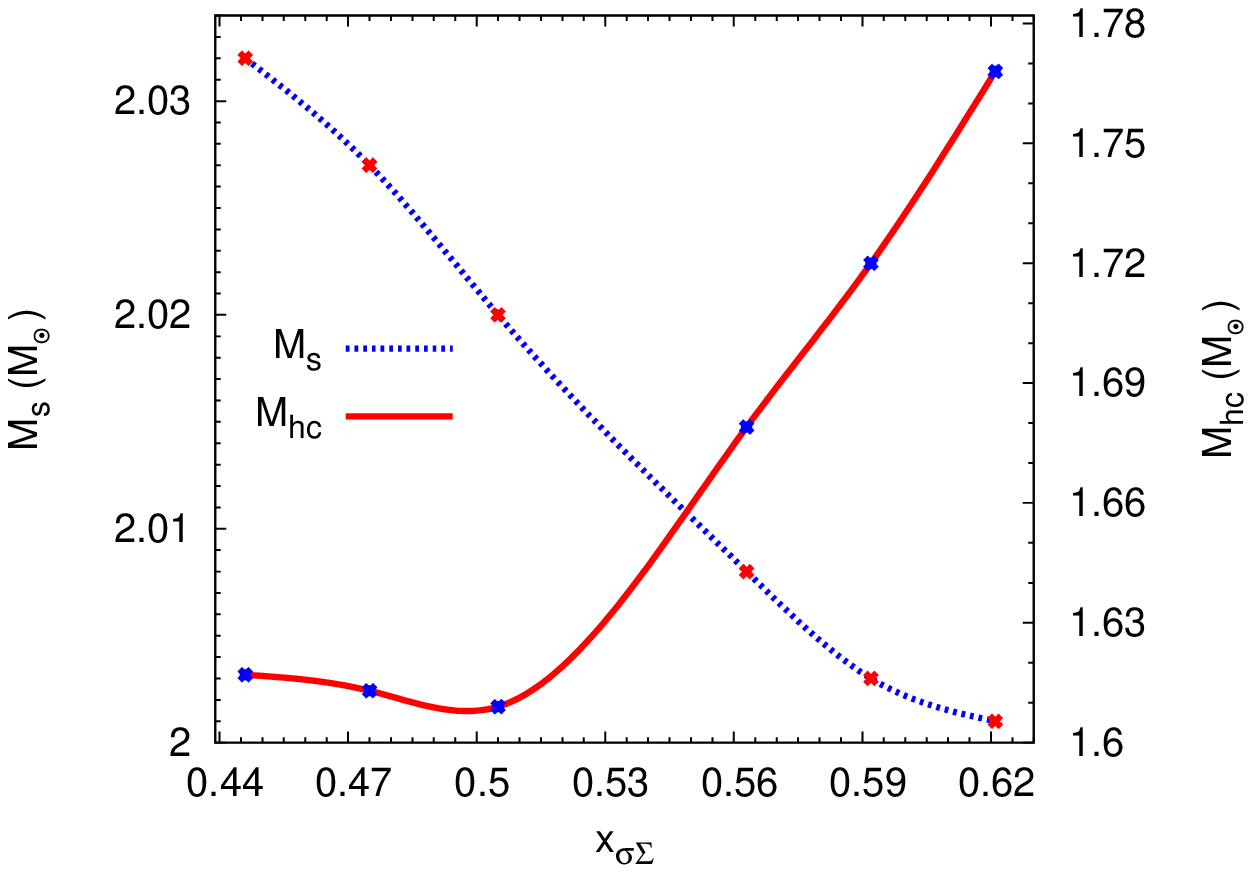} % eosTM1.eps: 1048592x1048592 pixel, 300dpi, 8878.08x8878.08 cm, bb=(atend)
\caption{(Color online) The dotted line shows the dependence of the maximum mass $M_{S}$ obtained for the given EoS on the value of parameter $x_{\Sigma \sigma}$ that expresses the dependence on the potential $U^{(N)}_{\Sigma}$. Solid line presented similar relations for the mass of the hyperon core $M_{hc}$.}
\label{fig:mhcmg}
\end{figure}

\begin{figure}
\includegraphics[clip,width=8.57cm]{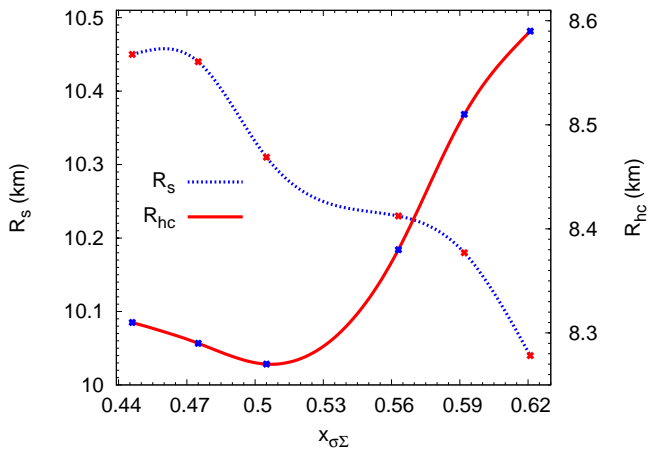} % eosTM1.eps: 1048592x1048592 pixel, 300dpi, 8878.08x8878.08 cm, bb=(atend)
\caption{(Color online) The dotted line shows the dependence of the radius of the maximum mass configurations $R_{S}$ obtained for the given EoS on the value of parameter $x_{\Sigma \sigma}$ that expresses the dependence on the potential $U^{(N)}_{\Sigma}$. Solid line presented similar relations for the radius of the hyperon core $R_{hc}$}
\label{fig:rhcrg}
\end{figure}

\begin{figure}
\includegraphics[clip,width=8.57cm]{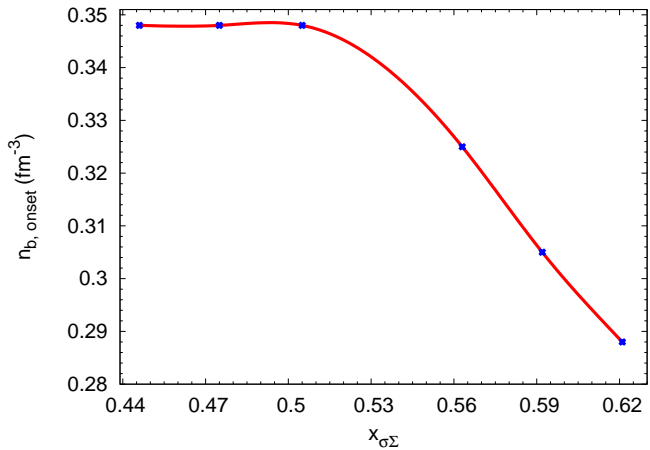} % eosTM1.eps: 1048592x1048592 pixel, 300dpi, 8878.08x8878.08 cm, bb=(atend)
\caption{(Color online) The dependence of the hyperon onset points on the value of the parameter $x_{\Sigma \sigma}$.}
\label{fig:onset}
\end{figure}

\begin{figure}
\includegraphics[clip,width=8.57cm]{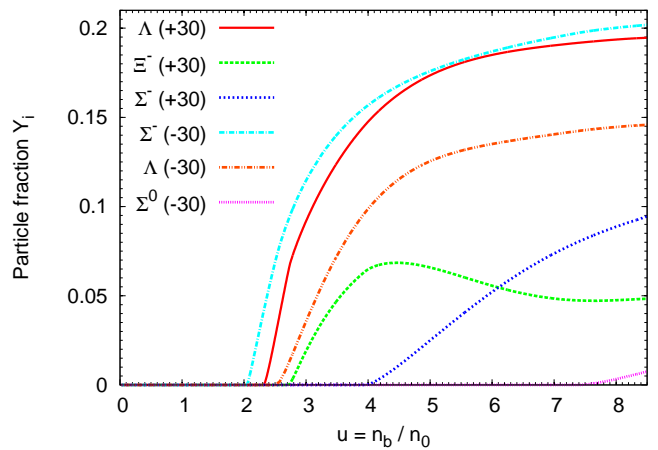} % eosTM1.eps: 1048592x1048592 pixel, 300dpi, 8878.08x8878.08 cm, bb=(atend)
\caption{(Color online) Comparison of the relative concentrations of hyperons calculated for the extended nonlinear model for the repulsive and attractive $U^{(N)}_{\Sigma}$ potential.}
\label{fig:u}
\end{figure}

%%%%%%%%%%%
There are still significant uncertainties  associated with the experimental data on the  hyperon-nucleus interactions.  
Thus it is reasonable to investigate  the effect of the hyperon-nucleus potential $U^{(N)}_{Y}$  on the obtained results \cite{Weissenborn2012}.
Particular attention was paid to the dependence of the EoS on the $\Sigma$-nucleus potential $U^{(\Sigma)}_{Y}$.
Detailed calculations were done for the selected values of the $U^{(N)}_{\Sigma}$ potential, 
assuming its both attractive and repulsive character.
Calculations performed for the extended nonlinear  model were resulted in a sequence of EoSs (Fig. \ref{fig:pot}).
 The stiffest one was obtained for the repulsive potential ($U^{(N)}_{\Sigma}= 30$ MeV). In order to study the influence of the $U^{(\Lambda)}_{\Lambda}$ potential a model with an exaggerated value of the potential $U^{(\Lambda)}_{\Lambda} = 1$ MeV was examined. The resulting change in the EoS is negligible. A similar conclusion can be drawn by examining the changes caused by the reduction in the potential $U^{(N)}_{\Xi}$. 
 Taking into account the value of the maximum mass obtained for a given EoS
the presented  results support the conclusion that
 the  value of the potential is not a factor that decisively influences the form of the EoS.
A similar conclusion can be obtained by analysing the mass-radius relation (Fig. \ref{fig:rm_pot}).
  The most promising results were obtained for the repulsive $U^{(N)}_{\Sigma}=-30$~MeV potential which within the considered  model leads to the highest value of the maximum mass.
Hyperons are distributed in the very inner part of a neutron star core therefore a schematic cross-section  shows a hyperon core in the inside of a strangeness-rich neutron star. Information concerning properties of the internal structure of a neutron star are summarised in Fig. \ref{fig:mhcmg} and \ref{fig:rhcrg}. These figures depicts the dependence of both  neutron stars and their  hyperon cores on the $ U^{(N)}_{\Sigma} $ potential and strictly speaking on the value of $x_{\Sigma \sigma}$ parameter and offer indications for interpreting the results of numerical calculations. Calculations were done for maximum mass configurations. The mass of the star increases when  $ U^{(N)}_{\Sigma} $ potential becomes more repulsive whereas the mass of the hyperon core decreases reaching a minimum value for the parameter $x_{\Sigma \sigma} \sim 0.51 $ (Fig. \ref{fig:mhcmg}). Similar behaviour has a radius of the maximum mass configuration. Results are presented in Fig.\ref{fig:rhcrg}. 
%%%%%%%%%%%
%%%%%%%%%%%
%%%%%%
\begin{table}
\caption{Neutron star maximum masses and corresponding radii calculated for the chosen values of the potentials $U^{(N)}_{\Sigma}$ and $U^{(\Lambda)}_{\Lambda}$. $M_{hc}$ and $R_{hc}$ denote the mass and radius of a hyperon core for a given maximum mass configuration, $n_{b,onset}$ denotes the hyperon onset point.}
\label{tab:core}
\begin{ruledtabular}
\begin{tabular}{rcccccc} 
\multicolumn{1}{c}{$U^{(N)}_{\Sigma}$} & $U^{(\Lambda)}_{\Lambda}$ & $M_{s}$ & $R_{s}$ & $M_{hc}$ & $R_{hc}$ & $n_{b,onset}$ \\ 
($MeV$) & ($MeV$) & ($M_{\odot}$) & ($km$) & ($M_{\odot}$) & ($km$) & ($fm^{-3}$) \\ \hline \hline
+30 & -1 & 2.036 & 10.47 & 1.620 & 8.32 & 0.348 \\ \hline
+30 & -5 & 2.032 & 10.45 & 1.617 & 8.31 & 0.348 \\ \hline
+20 & -5 & 2.027 & 10.44 & 1.613 & 8.29 & 0.348 \\ \hline
+10 & -5 & 2.020 & 10.31 & 1.609 & 8.27 & 0.348 \\ \hline
-10 & -5 & 2.008 & 10.23 & 1.679 & 8.38 & 0.325 \\ \hline
-20 & -5 & 2.003 & 10.18 & 1.720 & 8.51 & 0.305 \\ \hline
-30 & -5 & 2.001 & 10.04 & 1.768 & 8.59 & 0.288 \\
\end{tabular} 
\end{ruledtabular}
\end{table}
%%%%%%
%%%%%%%%%%%%%%%
%%%%%%%%%%%
%%%%%%%%%%%
One of the most important difference between the model constructed for the attractive $U^{(N)}_{\Sigma}=-30$ MeV potential and the one with the repulsive potential ($U^{(N)}_{\Sigma}=30$ MeV) is related to the chemical composition of the neutron star matter. The hyperon onset points depend on the character of the $U^{(N)}_{\Sigma}$ potential. Calculations performed on the basis of the extended nonlinear model indicate that reduction of the attractive $U^{(N)}_{\Sigma}$ potential shifts the hyperon onset points to lower densities, whereas the change of the repulsive potential leaves the hyperon onset points almost unchanged (Fig. \ref{fig:onset}).
In the case of attractive potential the first hyperons that appear in neutron star matter are $\Sigma^{-}$ hyperons and they are followed by $\Lambda$ hyperons. Such a scheme of the emergence of hyperons is changed in the nonlinear model with the repulsive   $U^{(N)}_{\Sigma}=-30$ MeV potential. The first hyperons that appear are $\Lambda$ hyperons and successively $\Xi^{-}$ and $\Sigma^{-}$ hyperons. The relative concentrations of hyperons calculated for both attractive and repulsive  $U^{(N)}_{\Sigma}$ potential are presented in Fig. \ref{fig:u}.  The results of numerical calculations obtained for different values of the potential  $U^{(N)}_{\Sigma}$ were  compiled in Table \ref{tab:core}. These results for a given EoS include the value of the maximum mass, the radius of the maximum mass configuration, the radius and mass of the hyperon core and the onset point of hyperons.
For comparison the model with the $U^{(\Lambda)}_{\Lambda}=1$ MeV was included.
%%%%%%%%%%%
%%%%%%%%%%%%%%%

%%%%%%%%%%%%%%%%%%%%%%%%%%%%%%%%%%%%%%%%%%%%%%%%%%%%%%%%%%%%%%%%%%%%%%
\section{Conclusions}

Recent observations of the binary millisecond pulsars give the value of neutron star mass about two solar masses. 
In general, theoretical models that describe strangeness-rich neutron star matter lead to much softer EoSs than those constructed for a core of a neutron star  composed of only  nucleons. This significant softening of the EoS causes a reduction in the maximum mass achievable in a given theoretical model.
This leads to an inconsistency between the  theoretical models that involve hyperons and the observation.
The solution to this problem, is the construction of a theoretical model of the matter of a neutron star  with hyperons that will give a high value of the maximum mass that demands a  stronger repulsive force in the strange sector  of the system.
The main issue that was addressed in this paper concerns  the influence of the nonlinear $\omega$, $\rho$ and $\phi $ vector mesons on the properties of asymmetric nuclear matter with a non-zero strangeness and consequently on the neutron star parameters.
A description of  neutron star matter given in this paper was done on the basis of a model inspired by the nonlinear realization of the chiral symmetry. 
Detailed analysis based on $SU(6)$ and $SU(3)$ symmetry of the behavior of the EoS for different relativistic models have been already done \cite{Weissenborn2012su3,Weissenborn2013,Miyatsu2013}. Whereas the analysis of the EoS of hypernuclear matter within a relativistic density functional theory with density-dependent couplings has been included in \cite{Colucci2013}.
An essential aspect of this model is  the diverse  vector meson sector which comprises  different vector meson couplings. 
As a result, a special class of EoSs adequate to characterise asymmetric nuclear matter with a non-zero strangeness was obtained. Particular EoSs are characterised  by the value of  parameter $\Lambda_{V}$, which determines the strength of the vector meson couplings. This very special form of the considered model gives  an EoS that is much stiffer than that obtained using the standard TM1-weak model.
Numerical solutions   permitted  the influence of the nonlinear vector meson couplings on the density dependence  of the scalar and vector meson fields to be investigated. This directly translates to the in-medium properties of baryons and mesons and leads to a considerable modification of the effective baryon and vector meson masses, especially in the strange sector of the model.

The nonlinear vector meson couplings modify both the asymmetry and strangeness content of the system and therefore lead to a  model with a
reduced strangeness and an enhanced asymmetry.
The stiffness of the EoS is directly related to the in-medium properties of the matter of a neutron star.
It depends on value of the effective baryon and meson masses, which modify the compressibility of the matter of a neutron star  and the range of interactions especially in the strange sector.
The results of the  analysis performed in the framework of the nonlinear model have shown  that the properties of neutron stars are significantly altered by the presence of hyperons.

The analysis of the dependence of neutron star  parameters on the strength of hyperon-nucleon interaction has been already done by Weissenborn et al. \cite{Weissenborn2012}. In this paper similar analysis has been performed and in this case the change of the value of hyperon-nucleon potential only slightly modifies the maximum mass of a neutron star, preferring a repulsive character of this potential. However, the changes of the hyperon-nucleon potential influence the parameters of the hyperon core of the neutron star.

The inclusion of hyperons does not soften the EoS; on the contrary, it leads to its considerable stiffening. The consequences for the parameters of neutron stars are straightforward and appear as the considerable growth of neutron star masses. 
 
In general, the hyperon fraction is reduced in comparison to  the linear models. The reduction of the hyperon population in the matter of a neutron star is related to the enhanced lepton concentrations. As an example the maximum mass configuration is presented. In this particular model the core of a neutron star reveal a hyperon inner core with the reduced $\Lambda$ and $\Xi^{-}$ hyperon concentrations but with the enhanced population of $\Sigma^{-}$ hyperons. \\

%%%%%%%%%%%%%%%%%%%%%%%%%%%%%%%%%%%%%%%%%%%%%%%%%%%%%%%%%%%%%%%%%%%%%%
\begin{acknowledgments}
M. Pienkos would like to acknowledge a scholarship from the SWIDER project co-finance by the European Social Fund.
\end{acknowledgments}

%%%%%%%%%%%%%%%%%%%%%%%%%%%%%%%%%%%%%%%%%%%%%%%%%%%%%%%%%%%%%%%%%%%%%%
%\nocite{*}
\bibliography{properties}% Produces the bibliography via BibTeX.
\end{document}